# ENHANCING DATA PRIVACY IN WIRELESS SENSOR NETWORKS: INVESTIGATING TECHNIQUES AND PROTOCOLS TO PROTECT PRIVACY OF DATA TRANSMITTED OVER WIRELESS SENSOR NETWORKS IN CRITICAL APPLICATIONS OF HEALTHCARE AND NATIONAL SECURITY

Akinsola Ahmed, Ejiofor Oluomachi, Akinde Abdullah, Njoku Tochukwu,

Department of Computer Science, Austin Peay State University, Clarksville USA.

*ABSTRACT*

*The article discusses the emergence of Wireless Sensor Networks (WSNs) as a groundbreaking technology in data processing and communication. It outlines how WSNs, composed of dispersed autonomous sensors, are utilized to monitor physical and environmental factors, transmitting data wirelessly for analysis. The article explores various applications of WSNs in healthcare, national security, emergency response, and infrastructure monitoring, highlighting their roles in enhancing patient care, public health surveillance, border security, disaster management, and military operations. Additionally, it examines the foundational concepts of data privacy in WSNs, focusing on encryption techniques, authentication mechanisms, anonymization techniques, and access control mechanisms. The article also addresses vulnerabilities, threats, and challenges related to data privacy in healthcare and national security contexts, emphasizing regulatory compliance, ethical considerations, and socio-economic factors. Furthermore, it introduces the Diffusion of Innovation Theory as a framework for understanding the adoption of privacy-enhancing technologies in WSNs. Finally, the article reviews empirical studies demonstrating the efficacy of security solutions in preserving data privacy in WSNs, offering insights into advancements in safeguarding sensitive information.*

## 1. INTRODUCTION

In the domains of data processing, communication systems, and sensor technologies, wireless sensor networks offer a new paradigm. Wireless sensor networks are composed of spatially scattered autonomous sensors that operate together to monitor physical or environmental parameters such as temperature, motion, pressure, humidity, and chemical concentrations (Kushala & Shylaja 2020). Tefera & Yang (2019) state that these sensors collect, process, and transmit data to a central location for analysis and decision-making via wireless communication. However, wireless sensor networks may be utilized for a variety of tasks in a number of sectors due to their affordability, scalability, and adaptability. There are more wireless sensor networks now than there were previously because to downsizing, wireless communication protocols, and sensor technology advancements. This has changed conventional methods for gathering and analyzing data and stimulated innovation (Wang et al., 2018).

Moreover, Ali et al. (2017) pointed out that wireless sensor networks are essential to advancing medical research, illness management, and patient care in the healthcare industry. Wireless sensor networks make real-time monitoring of patient's physiological characteristics, medication adherence, and vital signs possible, which seamlessly integrate sensors with medical equipment, wearable technology, and healthcare infrastructure (Lu et al., 2010). Ultimately, this ongoing observation improves patient outcomes and lowers healthcare costs by enabling prompt intervention, early identification of irregularities in health, and customized treatment plans. Furthermore, Abusaimeh et al. (2014) noted that wireless sensor networks allow for remote patient monitoring and telemedicine, allowing medical experts to assess patients' health and offer guidance or treatment from a distance. This is especially advantageous for patients in remote or underdeveloped locations, the elderly, and those with chronic diseases who require ongoing monitoring and care.

Additionally, wireless sensor networks are useful in medical research and public health surveillance. Wireless sensor networks allow researchers and policymakers to identify health disparities, track the spread of infectious diseases, and implement targeted interventions to improve population health outcomes by collecting and analyzing large amounts of health-related data, such as epidemiological trends, disease outbreaks, and environmental factors (Kushala & Shylaja 2020). However, Abusaimeh et al. (2014) stated that wireless sensor networks have the potential to transform healthcare delivery by improving patient monitoring, enabling remote care delivery, and facilitating data-driven decision-making, all of which advance the goals of precision medicine, population health management, and healthcare accessibility and equity.

Furthermore, wireless sensor networks aid in maintaining patient privacy. It guards against illegal access to patient data, which frequently results in confidentiality violations that violate patients' rights to privacy and put healthcare practitioners at risk of legal repercussions (Lu et al., 2010). Furthermore, wireless sensor networks support the maintenance of patient confidence in healthcare professionals. Sensitive information is handed to healthcare practitioners, who are supposed to handle it with care and discretion (Asaad 2021). Furthermore, Wang et al. (2018) pointed out that strict data privacy laws, such the Health Insurance Portability and Accountability Act (HIPAA) in the US, apply to healthcare institutions. If these rules are broken, there may be harsh consequences, such as fines and legal repercussions.

Likewise, Jan et al. (2019) observed that wireless sensor networks are vital instruments for monitoring, reconnaissance, and situational awareness in both civilian and military contexts when it comes to national security. Wireless sensor networks allow for continuous monitoring of human, environmental, and physical activity, and can be used to detect and deter security threats, illicit activities, and security breaches by placing sensors in strategic locations, border regions, and critical infrastructure (Lu et al., 2010). Additionally, wireless sensor networks in border surveillance and perimeter protection offer real-time monitoring of unlawful movements, suspicious activities, and border crossings, allowing security professionals and border patrol officers to react quickly to possible security breaches. It strengthens border security, stops smuggling of contraband, human trafficking, and illegal immigration, and protects the integrity of national borders (Tefera & Yang 2019).

Additionally, wireless sensor networks are essential for emergency response and catastrophe management. Wireless sensor networks provide early detection of man-made events like terrorist attacks or industrial accidents as well as natural disasters like earthquakes, floods, and wildfires by placing sensors in metropolitan centers, disaster-prone areas, and key infrastructure. This early identification improves public safety and lessens the effect of disasters on lives and property by facilitating quick response coordination, evacuation planning, and resource allocation (Chow et al., 2011). Additionally, wireless sensor networks are employed in military surveillance, information gathering, and environmental monitoring applications. Wireless sensor networks offer important insights for military operations, homeland protection, and strategic planning by gathering and evaluating data on topographical features, environmental conditions, and enemy movements (Asaad 2021).

Furthermore, vital infrastructure including power plants, transit systems, and government buildings is observed through the deployment of wireless sensor networks. Vulnerabilities in infrastructure systems may be revealed by breaches of data privacy in these networks, giving attackers the opportunity to do harm, interfere with operations, or launch cyberattacks (Bellare & Hoang 2015). Abusaimeh et al. (2014) pointed out that military troops in the battlefield depend on wireless sensor networks for surveillance, communication, and situational awareness. To ensure operational security, tactical advantage, and the accomplishment of military tasks, military intelligence data privacy must be protected. Still, the primary goal of the research is to assess how well wireless sensor networks protect patient privacy in vital national security and healthcare applications.

## 2. LITERATURE REVIEW

Sensitive data transmission across wireless sensor networks is more likely due to the growth of IoT (Internet of Things) devices, which include sensors implanted in a variety of objects and settings (Lightfoot et al., 2010). Furthermore, Dong et al. (2015) pointed out that there might be more serious repercussions from privacy breaches when wireless Sensor Networks are used more often in vital industries like national security and healthcare. To ensure the integrity, confidentiality, and security of data transferred across wireless sensor networks, it is critical to protect sensitive information from unwanted access, modification, or exploitation.

### 2.1. Conceptual Foundations of Data Privacy in Wireless Sensor Networks

Fundamentally, data privacy in wireless sensor networks refers to safeguarding private information sent over these networks from illegal access, disclosure, or alteration. It comprises operational data like sensor readings or telemetry data gathered from IoT devices, as well as personal data like medical records in healthcare applications or classified intelligence in national security situations (Mahmoud & Shen 2012). A number of elements, including as legal and regulatory frameworks, technical breakthroughs, ethical concerns, and socio-cultural norms, influence the conceptual underpinnings of data privacy in wireless sensor networks (Long et al., 2014). Nonetheless, confidentiality the guarantee that private information is kept secret and available to only those with permission is one of the fundamental ideas supporting data privacy in wireless sensor networks (Mutalemwa & Shin 2019). According to Chen et al. (2015), secrecy is necessary to uphold people's right to privacy, preserve confidence in data custodians, and stop sensitive information from being disclosed without authorization. In wireless sensor networks, methods like access control, encryption, and secure communication protocols are used to maintain confidentiality, guaranteeing that information is kept private even in the event that it is captured by hostile parties (Mahmoud & Shen 2012).

Integrity, which deals with the precision, dependability, and consistency of data sent across wireless sensor networks, is yet another important concept. According to Dong et al. (2015), maintaining data integrity is crucial for safeguarding the reliability of information and averting illegal alteration or tampering. To ensure the integrity of data transferred across wireless sensor networks, methods including digital signatures, hash functions, and data authentication procedures are employed. These techniques also serve to identify and prevent attempts to modify or change data while it is being communicated (Tefera & Yang 2019). Additionally, data privacy in wireless sensor networks is influenced by the principle of availability, which refers to the accessibility and usability of data by authorized parties when needed. Ensuring data availability is critical for supporting real-time decision-making, emergency response, and mission-critical operations in healthcare, national security, and other sectors. Techniques such as redundancy, fault tolerance, and disaster recovery mechanisms are employed to ensure continuous availability of data in Wireless sensor networks, mitigating the impact of network failures, cyber-attacks, or other disruptions (Chen et al., 2015).

Moreover, data privacy in wireless sensor networks is informed by legal and regulatory frameworks that govern the collection, storage, and processing of sensitive information. In healthcare, for example, regulations such as the Health Insurance Portability and Accountability Act (HIPAA) in the United States and the General Data Protection Regulation (GDPR) in the European Union impose stringent requirements for protecting patient privacy and ensuring the security of health-related data (Mahmoud & Shen 2012). Similarly, in national security contexts, laws and policies dictate the handling of classified information and impose restrictions on access to sensitive intelligence data (Tefera & Yang 2019). However, ethical considerations also play a significant role in shaping data privacy in wireless sensor networks, as decisions about the collection, use, and dissemination of sensitive information may have profound ethical implications. Ethical principles such as respect for privacy, autonomy, and informed consent guide the ethical design and implementation of Wireless sensor networks, emphasizing the importance of transparency, accountability, and user empowerment in protecting individuals' privacy rights (Mutalemwa & Shin 2019).

Furthermore, the conceptual underpinnings of data privacy in wireless sensor networks are shaped by sociocultural norms and public attitudes toward privacy, which also affect how sensitive information is handled in society and how privacy risks are perceived (Tefera & Yang 2019). Culturally sensitive methods to data privacy that respect varied viewpoints and values are necessary because cultural differences in attitudes toward privacy may have an influence on the design and implementation of privacy protection mechanisms in wireless sensor networks (Lightfoot et al., 2010). In summary, a variety of frameworks, ideas, and principles influence data privacy in wireless sensor networks by providing insight into the management, security, and protection of sensitive data in these systems.

## 2.2. Techniques and Protocols for Enhancing Data Privacy in Wireless Sensor Networks

Wireless Sensor Networks are increasingly pervasive in modern society, facilitating the collection, transmission, and processing of data in various applications, including healthcare, environmental monitoring, and smart cities (Hanumantha, et al., 2015). Odeh et al. (2015) noted that the widespread adoption of Wireless sensor networks also raises concerns about data privacy, as sensitive information transmitted over these networks may be vulnerable to interception, tampering, or unauthorized access. To address these concerns, a range of techniques and protocols have been developed to enhance data privacy in Wireless sensor networks. It includes:

- **Encryption Techniques**: Encryption is a fundamental technique for protecting data privacy in Wireless sensor networks by encoding sensitive information in a way that renders it unintelligible to unauthorized parties. Symmetric encryption algorithms, such as Advanced Encryption Standard (AES) and Data Encryption Standard (DES), use a single secret key to encrypt and decrypt data, ensuring confidentiality and integrity of transmitted information (Asaad et al., 2017). Asymmetric encryption algorithms, such as Rivest-Shamir-Adleman (RSA) and Elliptic Curve Cryptography (ECC), employ a pair of public and private keys for encryption and decryption, enabling secure communication between nodes without the need for a shared secret key. Encryption techniques can be applied at various levels of the communication stack in Wireless sensor networks, including link-layer encryption, network-layer encryption, and application-layer encryption, depending on the specific security requirements and constraints of the application (Bagaria 2015).

- **Authentication Mechanisms**: Authentication is essential for verifying the identity of communicating entities in Wireless sensor networks and preventing unauthorized access to sensitive data. Digital signatures, based on asymmetric cryptography, provide a mechanism for authenticating the origin and integrity of transmitted messages by appending a unique signature generated using the sender's private key (Chase & Shen 2015). Message authentication codes (MACs), derived from symmetric cryptography, enable nodes to verify the authenticity and integrity of messages using a shared secret key. Mutual authentication protocols, such as Transport Layer Security (TLS) and Secure Shell (SSH), establish bidirectional trust between communicating entities by exchanging digital certificates and verifying each other's identities. Authentication mechanisms play a crucial role in ensuring the trustworthiness of data transmitted over Wireless sensor networks and mitigating the risk of impersonation, replay attacks, and man-in-the-middle attacks (Bellare & Hoang 2015).

- **Anonymization Techniques**: Anonymization is a privacy-enhancing technique that aims to protect the identity and sensitive attributes of individuals in Wireless sensor networks by concealing or obfuscating personal information. Methods such as data masking, pseudonymization, and k-anonymity are commonly used to anonymize data before transmission, ensuring that sensitive information cannot be linked back to specific individuals (Chen et al., 2011). Data masking techniques, such as randomization and perturbation, modify the values of sensitive attributes to prevent identification of individuals while preserving the statistical properties of the data. Pseudonymization replaces identifying attributes with pseudonyms or aliases to anonymize data without losing its utility for analysis or processing. k-anonymity ensures that each record in a dataset is indistinguishable from at least k-1 other records, reducing the risk of re-identification through record linkage or inference attacks. Anonymization techniques can be applied at the data collection stage, the data transmission stage, or the data storage stage in Wireless sensor networks, depending on the privacy requirements and constraints of the application (Abdalla et al., 2015).

- **Access Control Mechanisms**: Access control is essential for regulating the permissions and privileges of users or devices accessing sensitive data in Wireless sensor networks and enforcing data privacy policies. Role-based access control (RBAC) assigns users or devices to roles with specific privileges and permissions based on their organizational roles or responsibilities (Huang et al., 2017). Discretionary access control (DAC) allows data owners to specify access control policies for individual resources or data objects, granting or denying access based on user identities or attributes. Mandatory access control (MAC) enforces access control policies mandated by a central authority, such as a security policy administrator or a trusted computing base. Attribute-based access control (ABAC) uses attributes and policies to dynamically determine access decisions based on contextual information, such as user attributes, resource attributes, and environmental conditions. Access control mechanisms can be implemented at various levels of the communication stack in Wireless sensor networks, including node-level access control, network-level access control, and application-level access control, to ensure that only authorized users or devices can access sensitive data and resources (Hanumantha, et al., 2015).

- **Secure Communication Protocols**: Secure communication protocols are essential for ensuring the confidentiality, integrity, and authenticity of data transmitted over Wireless sensor networks and protecting against eavesdropping, tampering, and spoofing attacks. Protocols such as Secure Sockets Layer (SSL), Transport Layer Security (TLS), and Datagram Transport Layer Security (DTLS) provide end-to-end encryption, authentication, and integrity protection for data exchanged between communicating entities in Wireless sensor networks. Lightweight protocols, such as TinySec and µTESLA, are specifically designed for resource-constrained Wireless sensor networks, providing efficient security mechanisms suitable for low-power devices with limited computational capabilities. Secure routing protocols, such as Secure Multipath Routing (SMR) and Authenticated Routing for Ad Hoc Networks (ARAN), establish secure communication paths between nodes in Wireless sensor networks, ensuring that data is routed securely and reliably through the network. Secure communication protocols play a critical role in ensuring the security and privacy of data transmitted over Wireless sensor networks, enabling secure communication between nodes, gateways, and backend systems while minimizing the risk of security breaches and unauthorized access (Martins & Sousa 2015).

## 2.3. Vulnerabilities, Threats, and Challenges Related to Data Privacy Concerns in Wireless Sensor Networks in Healthcare and National Security

Wireless Sensor Networks have revolutionized various sectors, including healthcare and national security, by enabling real-time monitoring, data collection, and analysis. However, the widespread deployment of Wireless sensor networks has also introduced new vulnerabilities, threats, and challenges related to data privacy. Vulnerabilities, threats, and challenges related to data privacy concerns in Wireless Sensor Networks pose significant risks and implications for healthcare and national security applications (Al-Omar et al., 2012).

Vulnerabilities in wireless sensor networks refer to weaknesses or flaws in the network architecture, protocols, or devices that can be exploited by attackers to compromise data privacy. One common vulnerability is the lack of encryption and authentication mechanisms in wireless sensor networks, which can expose sensitive information to eavesdropping, interception, or tampering (Asaad 2020). Inadequate access control mechanisms and weak authentication mechanisms also pose significant vulnerabilities, allowing unauthorized users or devices to gain access to sensitive data or resources. Moreover, the resource-constrained nature of wireless sensor networks devices, such as limited processing power, memory, and battery life, introduces vulnerabilities that can be exploited by attackers to launch denial-of-service attacks, physical attacks, or side-channel attacks (Marqas et al., 2022).

Additionally, threats to data privacy in wireless sensor networks encompass various malicious activities and attacks aimed at exploiting vulnerabilities and compromising the confidentiality, integrity, or availability of sensitive information. One common threat is eavesdropping, where attackers intercept and monitor communication between nodes in wireless sensor networks to obtain sensitive data, such as patient health records in healthcare applications or classified intelligence in national security contexts (Kumari et al., 2019). Another threat is data tampering, where attackers modify or manipulate data transmitted over wireless sensor networks to disrupt operations, falsify information, or deceive users. Denial-of-service (DoS) attacks, such as jamming attacks or node exhaustion attacks, pose a threat to data availability by disrupting communication or rendering wireless sensor networks nodes inoperable. Furthermore, insider threats, such as rogue employees or compromised devices, can pose significant risks to data privacy by exploiting their privileged access or knowledge of the network (Asaad & Abdulnabi 2022).

## 2.4. Theoretical Review

The study employed the Diffusion of Innovation Theory to explain data privacy concerns in Wireless Sensor Networks (WSNs) within healthcare and national security.

### 2.4.1. Diffusion of InnovationTheory

Diffusion of Innovation Theory was proposed by Everett Rogers in 1962. The theory seeks to explain how and why new ideas, products, or technologies spread through societies or organizations over time (Rogers 1962). Rogers, a communication scholar and sociologist, conducted extensive research on the diffusion process across various domains, including agriculture, public health, and technology adoption. His seminal work, "Diffusion of Innovations," outlines the key principles and stages of the diffusion process, shedding light on the factors influencing the adoption and acceptance of innovations (Rogers et al., 1971).

Also, the theory posits that the adoption of innovations follows a predictable pattern characterized by the diffusion curve, which depicts the cumulative percentage of adopters over time. According to Rogers, innovators are the first individuals or organizations to adopt new innovations, followed by early adopters, early majority, late majority, and laggards. Each adopter category represents a distinct segment of the population with varying degrees of innovativeness, risk aversion, and susceptibility to social influence (Barnett 1953).

Additionally, the application of Diffusion of Innovation Theory to enhancing data privacy in Wireless Sensor Networks (WSNs) involves understanding how innovations in privacy-enhancing technologies and protocols are adopted and diffused among users and organizations within the WSN ecosystem. At the forefront of the diffusion process are innovators, who are typically early adopters of new privacy-enhancing technologies or protocols in Wireless sensor networks. These innovators are often technologically savvy individuals, organizations, or research institutions that recognize the potential benefits of adopting innovative solutions to address data privacy concerns in wireless sensor networks. For example, research institutions may develop and test novel encryption algorithms or authentication mechanisms designed to protect sensitive data transmitted over Wireless sensor networks, positioning themselves as innovators in the field (O'Connor 2007).

Furthermore, following the innovators are early adopters, who are influential stakeholders or opinion leaders within the WSN community. Early adopters may include government agencies, industry associations, or leading enterprises that are motivated to adopt privacy-enhancing technologies or protocols to gain a competitive edge, enhance security, or comply with regulatory requirements. These early adopters play a crucial role in legitimizing and diffusing innovations in data privacy across the broader WSN ecosystem, serving as role models and sources of information for other users and organizations. As innovations in data privacy gain momentum, they begin to penetrate the early majority segment of WSN users and organizations. The early majority consists of pragmatists who are motivated by the perceived benefits and practical applications of privacy-enhancing technologies or protocols. For example, healthcare providers may adopt encryption techniques or access control mechanisms to safeguard patient health records transmitted over Wireless sensor networks, driven by concerns about data breaches and regulatory compliance (Lyytinen & Damsgaard 2001).

Additionally, the late majority and laggards segments of WSN users and organizations begin to adopt privacy-enhancing innovations, albeit at a slower pace and with greater resistance to change. The late majority may include smaller enterprises, non-profit organizations, or public institutions that follow the lead of early adopters and pragmatists, while laggards may comprise individuals or organizations that are hesitant or skeptical about adopting new technologies or protocols due to inertia, skepticism, or resource constraints. Overall, applying the Diffusion of Innovation Theory to enhancing data privacy in Wireless Sensor Networks involves understanding the dynamics of innovation adoption and diffusion among different segments of users and organizations within the WSN ecosystem. By identifying and targeting early adopters, leveraging social networks, and addressing user perceptions and contextual factors, organizations can accelerate the adoption and diffusion of privacy-enhancing technologies and protocols in Wireless sensor networks, ultimately promoting trust, security, and privacy in the digital age (Ismail 2006).

## 3. METHODOLOGY

In keeping with the merging of the deductive and inductive research methodologies, this study utilises a single-method research design. This is a quantitative method of data collection and analysis. The population of this study were professionals with expertise in wireless sensor networks, cybersecurity, and privacy, especially those working in critical applications of healthcare and national security. The decision to gather relevant information from this group was dictated by their experience implementing and managing wireless sensor networks in these critical applications. As a result of this, 865 professionals participated in the survey exercise, as it was necessary to collect relevant and valid information to achieve the study's objective.

The study adopted an online questionnaire, with a link to the form sent to groups and associations of professionals with expertise in wireless sensor networks, cybersecurity, and privacy. The nature of the collected data is primary, given that direct opinions of these professionals were gotten directly from the source as this information has not been saved in any database prior to seeking their perception of techniques and protocols.

The study adopted descriptive statistics and regression analysis to examine trends, patterns, and the effect of the protocols and techniques on healthcare and national security as it related to data protection and privacy. The decision to adopt regression analysis is to identify the protocols and techniques that significantly influence data privacy and protection.

### 3.1. Model Specification

The model for the study encompasses the postulated techniques and protocols for data privacy, being the independent variables, while the perception of the participants on healthcare and national security represents the dependent variable. Functionally, the model is specified as follows:

HNS = f(ENC, AUT, ACC, ANY, SRP, PHS, SDA, IDPS, SKY, DAM) ………… Eqn 1

Equation 1 above can be linearly written as follows;

$HNS_i = \beta_0 + \beta_1 ENC_i + \beta_2 AUT_i + \beta_3 ACC_i + \beta_4 ANY_i + \beta_5 SRP_i + \beta_6 PHS_i + \beta_7 SDA_i + \beta_8 IDPS_i + \beta_9 SKM_i + \beta_{10} DAM_i + \mu_i$ ………………………………………………………..Eqn II

Where;

HNS = Healthcare and National Security, ENC = Encryption, AUT = Authentication, ACC = Access Control, ANY = Anonymisation, SRP = Secure Routing Protocols, PHS = Physical

Security, SDA = Secure Data Aggregation, IDPS = Intrusion Detection and Prevention System, SKM = Secure Key Management, DAM = Data Masking

## 4. RESULT AND DISCUSSION

This section encapsulates the exposition, examination, and dialogues surrounding the dataset employed in this study. It elucidates and extrapolates the outcomes of the econometric modeling concerning the seminar's research inquiries and goals. Here, the descriptive essence of the variables is delineated alongside the outcomes of diverse determinants across both the overall and subset datasets.

### 4.1. Descriptive Analysis

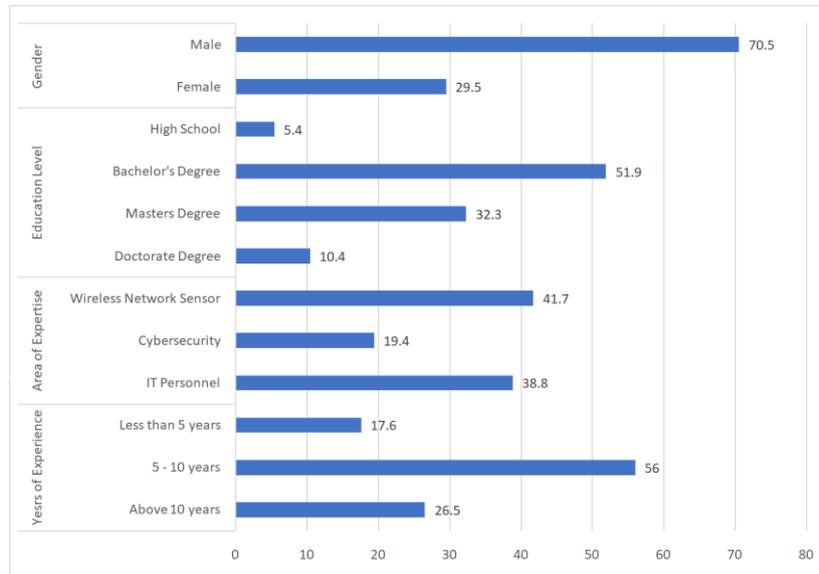

Figure 1: Socio-Demographic Characteristics of the Participants (%)

The figure above presents the responses obtained in relation to the socio-demographic characteristics of the participants. It reveals a demographic composition that is predominantly male, with 70.5% of the respondents identifying as male and 29.5% as female. In terms of education, a majority of the respondents hold a Bachelor's degree (51.9%), followed by those with a Master's degree (32.3%), while only a small percentage have a Doctorate's degree (10.4%) or a High School education (5.4%). Regarding expertise, the survey indicates that 41.7% of the respondents specialize in Wireless Network Sensors, followed by 38.8% in IT Personnel, and 19.4% in Cybersecurity. In terms of experience, a significant portion of the respondents (56.0%) have 5-10 years of experience, while 26.5% have more than 10 years of experience, and 17.6% have less than 5 years of experience. The mean age of the respondents is 38.91 years, with a minimum age of 27 years and a maximum age of 50 years, indicating a relatively spread-out age distribution. Overall, the data provides valuable insights into the demographic and professional characteristics of the participants.

Table 4.1: Perception on Healthcare and National Security

| Questions | Mean | Std. Dev. |
|---|---|---|
| Applying different protocols and techniques to data guarantee privacy. | 3.55 | 1.116 |
| I believe that the protection of health data is essential for national security. | 3.76 | .932 |
| There are adequate measures in place to protect health and national data. | 3.47 | 1.166 |
| I am aware of the potential risks associated with the transmission of health data over wireless networks. | 3.97 | .862 |
| I believe that improving the security of healthcare data can enhance the overall quality of healthcare services. | 4.23 | .716 |

Table 4.1 presents the views of the participants regarding healthcare and national security. The perception of professionals in cybersecurity, IT personnel, and wireless network sensor fields regarding healthcare and national security is moderately positive overall, as indicated by the mean scores. Participants generally agree that applying different protocols and techniques can guarantee data privacy (mean = 3.55). There is a stronger agreement that the protection of health data is essential for national security (mean = 3.76). However, respondents are less convinced about the adequacy of measures in place to protect health and national data (mean = 3.47). On the other hand, participants show a high level of awareness regarding the potential risks associated with the transmission of health data over wireless networks (mean = 3.97). This suggests a recognition of the challenges and vulnerabilities in current data transmission practices. Moreover, there is a strong belief that improving the security of healthcare data can enhance the overall quality of healthcare services (mean = 4.23).

These results reveal a sophisticated awareness among professionals on the relevance of data privacy in healthcare and its implications for national security. While there is a universal agreement on the need for stronger security measures, there are still questions about the adequacy of present processes.

International Journal of Network Security & Its Applications (IJNSA) Vol.16, No.2, March 2024

Table 4.2: Data Privacy Techniques and Protocols

| Questions | Mean | Std. Dev. |
|---|---|---|
| Encryption methods are essential in ensuring the privacy of data transmitted over wireless sensor networks | 3.62 | .969 |
| Authentication mechanisms contribute to securing data in critical applications of healthcare and national security | 4.16 | .598 |
| The effectiveness of access control techniques contributes to prevention of unauthorized access to sensitive data | 4.10 | .814 |
| I consider data anonymization techniques relevant in protecting the privacy of individuals in wireless sensor networks | 4.08 | .810 |
| I trust data masking techniques in concealing the meaning of data to unauthorized parties | 4.03 | .832 |
| I am confident that intrusion detection systems are effective for identifying and responding to unauthorized access attempts | 3.54 | 1.076 |
| I find secure routing protocols effective in ensuring that data is transmitted securely within wireless sensor networks | 3.65 | 1.100 |
| I am concerned about the potential risks posed by data aggregation techniques in compromising the privacy of data | 4.22 | .669 |
| I am confident that secure data storage mechanisms contribute to the overall security of data in wireless sensor networks | 4.06 | .849 |
| The current state of physical security protocols used to protect data in wireless sensor networks is satisfactory | 4.09 | .779 |

From Table 4.2, the perception of professionals regarding data privacy techniques and protocols in wireless sensor networks for critical applications of healthcare and national security is generally positive, with mean scores above the midpoint of 3. Participants strongly agree that authentication mechanisms contribute to securing data (mean = 4.16) and that access control techniques are effective in preventing unauthorized access (mean = 4.10). There is also a high level of agreement on the relevance of data anonymization techniques (mean = 4.08) and data masking techniques (mean = 4.03) in protecting privacy. However, there is some concern about the effectiveness of intrusion detection systems (mean = 3.54) and secure routing protocols (mean = 3.65), indicating room for improvement in these areas. Participants are particularly concerned about the potential risks posed by data aggregation techniques (mean = 4.22), highlighting the need for careful consideration of privacy implications when using such techniques. Overall, there is confidence in the effectiveness of secure data storage mechanisms (mean = 4.06) and physical security protocols (mean = 4.09) in ensuring the security of data in wireless sensor networks.

Table 4.3: Consequences of Data Privacy Breaches

| Questions | Mean | Std. Dev. |
|---|---|---|
| I believe that privacy breaches in wireless sensor networks could lead to unauthorized access to sensitive data | 3.97 | .843 |
| Loss of confidentiality arises from privacy breaches | 4.07 | .801 |
| Privacy breaches could result in the manipulation or tampering of healthcare or national security data | 4.02 | .880 |
| Privacy breaches disrupt critical operations, which can be harmful to an economy | 3.34 | 1.227 |
| People tends to lose their trust in information stored in the database when breaches occur | 3.71 | 1.062 |

Based on the outcome in Table 4.3, there are notable consequences that may arise as a result of data privacy breaches. There is a high level of agreement that privacy breaches could lead to unauthorized access to sensitive data (mean = 3.97) and loss of confidentiality (mean = 4.07). Participants also believe that privacy breaches could result in the manipulation or tampering of healthcare or national security data (mean = 4.02) and disrupt critical operations, which can be harmful to an economy (mean = 3.34). While there is agreement that people tend to lose their trust in information stored in the database when breaches occur (mean = 3.71), the mean score is slightly lower compared to other statements. Summarily, the findings suggest that professionals are aware of the serious implications of privacy breaches in wireless sensor networks and emphasize the importance of implementing robust security measures to mitigate these risks.

## 4.2. Inferential Analysis

This section examines the effect of the techniques and protocols on healthcare and national security as it relates to data privacy using regression method.

Table 4.4: Model Summary

| Model | R | R Square | Adjusted R Square | Std. Error of the Estimate |
|---|---|---|---|---|
| 1 | .232[a] | .054 | .043 | 1.092 |
| a. Predictors: (Constant), ENC, AUT, ACC, ANY, PHS, IDPS, SRP, SDA, SKM, DAM | | | | |

Table 4.4 presents the model summary, which indicates that the R-squared value is 0.054. The implication of this is that 5.4% variation in healthcare and national security is explained by the independent variables included in the study. The R statistic of 0.232, shows there is a weak correlation among the variables.

Table 4.5: ANOVA[a]

| Model | | Sum of Squares | Df | Mean Square | F | Sig. |
|---|---|---|---|---|---|---|
| 1 | Regression | 58.101 | 10 | 5.810 | 4.876 | .000[b] |
| | Residual | 1017.541 | 854 | 1.192 | | |
| | Total | 1075.642 | 864 | | | |
| a. Dependent Variable: HNS. | | | | | | |
| b. Predictors: (Constant), ENC, AUT, ACC, ANY, PHS, IDPS, SRP, SDA, SKM, DAM | | | | | | |

The ANOVA table summarizes the analysis of variance for the regression model. The table shows that the regression model is statistically significant, as indicated by the F-statistic of 4.876 and a corresponding p-value of .000 ($p < .001$). This suggests that all the independent variables (ENC, AUT, ACC, ANY, PHS, IDPS, SRP, SDA, SKM, and DAM) are jointly significant in predicting healthcare and national security. The regression sum of squares (58.101) represents the variation in healthcare and national security explained by the regression model, while the residual sum of squares (1017.541) represents the unexplained variation. Overall, the model accounts for a significant amount of the variance in healthcare and national security in terms of data privacy.

Table 4.6: Regression Analysis

| Coefficients | | | | | | |
|---|---|---|---|---|---|---|
| Model | | Unstandardized Coefficients | | Standardized Coefficients | t | Sig. |
| | | B | Std. Error | Beta | | |
| | (Constant) | 2.504 | 0.609 | | 4.110 | 0.000 |
| | ENC | 0.130 | 0.039 | 0.113 | 3.318 | 0.001 |
| | AUT | 0.032 | 0.063 | 0.017 | 0.505 | 0.614 |
| | ACC | 0.127 | 0.047 | 0.093 | 2.716 | 0.007 |
| | ANY | -0.019 | 0.047 | -0.014 | -0.406 | 0.685 |
| | PHS | -0.053 | 0.050 | -0.037 | -1.051 | 0.293 |
| | IDPS | 0.003 | 0.036 | 0.002 | 0.071 | 0.944 |
| | SRP | 0.183 | 0.035 | 0.180 | 5.218 | 0.000 |
| | SDA | -0.046 | 0.056 | -0.028 | -0.824 | 0.410 |
| | SKM | 0.010 | 0.044 | 0.008 | 0.235 | 0.815 |
| | DAM | -0.075 | 0.045 | -0.056 | -1.653 | 0.099 |
| a. Dependent Variable: Healthcare and National Security | | | | | | |

Explanatory Notes: ENC = Encryption, AUT = Authentication, ACC = Access Control, ANY = Anonymisation, SRP = Secure Routing Protocols, PHS = Physical Security, SDA = Secure Data Aggregation, IDPS = Intrusion Detection and Prevention System, SKM = Secure Key Management, DAM = Data Masking

The regression analysis aimed to examine the relationship between various data privacy techniques and protocols and their impact on Healthcare and National Security in wireless sensor networks.

Encryption (ENC) demonstrated a significant positive effect on Healthcare and National Security (B = 0.130, p = 0.001). This implies that the use of encryption methods is essential in ensuring the privacy of data transmitted over wireless sensor networks, thereby contributing to the overall security of healthcare and national security data. A percent point increase in encryption will bring about increase in healthcare and national security by 0.130 percent points.

Access Control (ACC) also showed a significant positive effect on Healthcare and National Security (B = 0.127, p = 0.007). This suggests that implementing effective access control mechanisms contributes to preventing unauthorized access to sensitive data, further enhancing the security of healthcare and national security information. That is, a percent point increase in implementation of access control will lead to increase in healthcare and national security by 0.127 percent points.

Additionally, Secure Routing Protocols (SRP) exhibited the highest positive effect on Healthcare and National Security (B = 0.183, p = 0.000). This indicates that using secure routing protocols is effective in ensuring that data is transmitted securely within wireless sensor networks, thereby contributing significantly to the overall security of healthcare and national security data. This alos implies that a percent point increase in the use of secure routing protocol will lead to increase in healthcare and national data security.

However, other variables such as Authentication (AUT), Anonymisation (ANY), Physical Security (PHS), Intrusion Detection and Prevention System (IDPS), Secure Data Aggregation (SDA), Secure Key Management (SKM), and Data Masking (DAM) did not show significant effects on Healthcare and National Data Security based on their p-values (p > 0.05).

## 5. CONCLUSION AND RECOMMENDATION

The study demonstrates that encryption methods play a pivotal role in ensuring the privacy of data transmitted over wireless sensor networks. The positive relationship between encryption and the security of healthcare and national security data suggests that implementing robust encryption mechanisms is essential. This finding underscores the importance of encryption as a fundamental tool in protecting sensitive information in critical applications. The study highlights the importance of access control mechanisms in preventing unauthorized access to sensitive data. The significant positive effect of access control on healthcare and national security indicates that implementing effective access control measures is crucial for enhancing data security. This finding emphasizes the need for organizations to implement strong access control policies and technologies to protect sensitive information. Furthermore, the study reveals that secure routing protocols are effective in ensuring the secure transmission of data within wireless sensor networks. The significant positive relationship between secure routing protocols and healthcare and national security suggests that implementing secure routing protocols is essential for protecting data during transmission. This finding underscores the importance of secure routing protocols in maintaining the integrity and confidentiality of healthcare and national security data.

## 5.1. Recommendations

Overall, the findings of this study have several implications for policymakers, practitioners, and researchers.

1. Policymakers should prioritize the implementation of encryption methods, access control mechanisms, and secure routing protocols to enhance the security of healthcare and national security data in wireless sensor networks. Additionally, organizations should invest in robust encryption technologies and access control systems to protect sensitive information from unauthorized access.
2. Practitioners should be aware of the importance of data privacy techniques and protocols in maintaining the security of healthcare and national security data. They should continuously update their knowledge and skills in implementing these techniques to mitigate potential risks and vulnerabilities.

Additionally, researchers should continue to explore and develop new data privacy techniques and protocols to address emerging threats and challenges in healthcare and national security data security. The findings of this study provide valuable insights into the importance of data privacy techniques and protocols in protecting healthcare and national security data in wireless sensor networks. By implementing robust protocols, organizations can enhance the security of sensitive information and mitigate potential risks and vulnerabilities.

**AUTHORS**

**Akinsola Ahmed** is a graduate student at Austin Peay State University in Clarksville, Tennessee, pursuing a Master of Science in Computer Science. He is passionate about information security and AI, and he believes these technologies have the potential to make the world a safer and more equitable place as such he's on a journey to leave an indelible mark in the world of technology, inspiring the next generation of innovators.

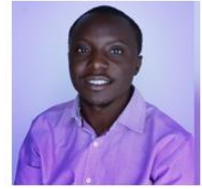

**Tochukwu Kennedy Njoku** is a highly motivated analyst with expertise in quantitative analysis, IT, research, and data-drivendecision-making. Currently, he is pursuing a Master of Science (MSc) degree in Computer Science and Quantitative Methods at Austin Peay State University, furthering his expertise in the ever-evolving field of technology. He is passionate about utilizing data to foster innovation and enhance solutions positively on a global scale.

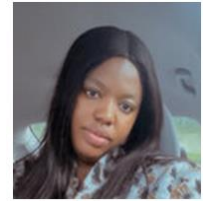

**Oluomachi Ejiofor** is a graduate student at Austin Peay State University studying Computer Science, specializing in Information Assurance and Security. She is passionate about cybersecurity, focusing on detecting threats, analyzing risks, and developing secure software. With a proactive approach and a dedication to continuous learning, she aims to contribute to the cybersecurity field and help organizations stay protected in today's digital landscape.

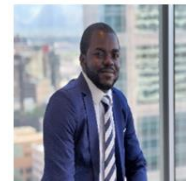

**Abdullah Akinde** is a skilled IT professional with a strong educational background and vast experience in the field. Earning his master's degree in computer science from Austin Peay State University after a bachelor's degree in systems engineering from the University of Lagos, giving him a solid foundation in both academic and practical skills. His certifications in AWS and Azure indicate his competence in cloud computing platforms, allowing him to harness cutting-edge technology to address complicated challenges. Abdullah was previously at Fidelity Bank plc, where he focused on data and application security vulnerability analysis.

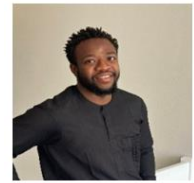